\documentclass{webofc}
\usepackage{subcaption}
\usepackage{hyperref}
\usepackage[font=small,labelfont=bf,textfont=normal,singlelinecheck=false]{caption}
\usepackage{booktabs}
\RequirePackage{orcidlink} 

\newenvironment{tightitemize}
  {\begin{itemize}
     \setlength{\leftmargin}{2.5em}
     \setlength{\itemindent}{2.5em}
     \setlength{\itemsep}{0pt}
     \setlength{\parskip}{0pt}
     \setlength{\parsep}{0pt}
     \setlength{\topsep}{0pt}}
  {\end{itemize}}

\begin{document}
\title{WLCG Mini-Capability Challenge: Host Tuning to Improve WAN Data Transfers}
%
%

\author{\firstname{Andrew Malone} \lastname{Melo}\inst{1}\orcidlink{0000-0003-3473-8858} \and
        \firstname{Carlos Fernando} \lastname{Gamboa}\inst{4} \orcidlink{0000-0001-5395-1588} \and
        \firstname{Diego} \lastname{Davila Foyo}\inst{3} \orcidlink{0000-0002-8664-5154}\and
        \firstname{Eduardo} \lastname{Bach}\inst{5} \orcidlink{0000-0002-8152-357X}\and
        \firstname{Eli} \lastname{Dart}\inst{6} \orcidlink{0000-0002-8229-5433}\and
        \firstname{Shawn} \lastname{McKee}\inst{2} \orcidlink{0000-0002-4551-4502}\and
        \firstname{Garhan} \lastname{Attebury}\inst{7} \orcidlink{0000-0003-0032-3638}\and
        \firstname{Philippe} \lastname{Laurens}\inst{8} \orcidlink{0000-0002-6203-0323}\and
        \firstname{Wendy} \lastname{Wu}\inst{2} \orcidlink{0000-0003-1128-884X}\and
        \firstname{Syed Asif} \lastname{Shah}\inst{9} \and
        \firstname{Hiro} \lastname{Ito}\inst{4} \orcidlink{0009-0002-7304-0831}\\
        }

\institute{ Vanderbilt University, Nashville, TN, USA
\and
            Physics Department, University of Michigan, Ann Arbor, MI, USA
\and
            University of California, San Diego, CA, USA
\and
            Brookhaven National Laboratory, Upton, NY, USA
\and
            University of Massachusetts, Amherst, MA, USA
\and
            Energy Sciences Network, Berkeley, CA, USA
\and
            University of Nebraska, Lincoln, NE, USA
\and
            Michigan State University, East Lansing, MI, USA
\and
            Fermi National Accelerator Laboratory, Batavia, IL, USA
          }

\abstract{%
High-throughput data movement is a defining requirement of the WLCG computing
model, yet end-to-end performance is often constrained less by the network backbone
than by host-level configuration. We present the results of a WLCG
mini-capability challenge focused on host optimization at ATLAS and CMS sites
(FNAL, UCSD, UNL, BNL, AGLT2, MWT2, and Vanderbilt), using EL8/9 hosts with 25\,Gbps or higher NICs. NET2 participated in a subsequent trans-Atlantic test. The methodology combines ESnet Fasterdata guidance with the OSG-hosted \texttt{fasterdata-tuning.sh} framework, applying TCP, queue, offload, and ring-buffer tuning with save and restore support for safe baseline comparisons. Results show that tuning benefit
depends critically on the local bottleneck. Where DTNs and storage were both
capable (UNL), FTS throughput rose from 60 to 90\,Gbps. Where storage was the
binding constraint (UCSD), iperf3 throughput improved from 20 to
90\,Gbps while FTS was unchanged. Where the NIC was the binding constraint (Vanderbilt, 25\,G NIC), the tuning profile actively reduced throughput, motivating a review of per-site
applicability. The challenge also surfaced and
resolved a dCache proxying misconfiguration at AGLT2, demonstrating that
mini-challenges serve a diagnostic function beyond raw performance measurement.
}
\maketitle

\section{Introduction}
\label{intro}
Data movement is a critical component of the LHC computing model\cite{wlcg-model}. Data recorded in the detectors are processed, distributed, and analyzed in facilities distributed globally across the WLCG. These workflows rely on sustained high-rate transfers between compute and storage systems, frequently across large geographic distances and through shared research and education network infrastructure.

This effort was organized as a WLCG mini-capability challenge: a targeted, ad hoc exercise that complements the larger WLCG Data Challenges conducted roughly every two years\cite{DC24}. Mini-challenges focus on a specific operational question with a small set of willing sites, enabling rapid iteration without the coordination overhead of a full-scale event. The February 2026 host-tuning challenge was one such exercise, spanning two days of coordinated testing across USATLAS and USCMS sites.

From a site perspective, the end-to-end throughput of a transfer often depends not only on the WAN path but also on the behavior of the host at each endpoint. Default Linux and storage settings may be adequate for general-purpose services, but they are rarely optimized for long-distance, high-bandwidth transfers that dominate HEP workflows. TCP buffer sizing, queue lengths, interface ring buffers, offload settings, storage I/O parameters, and CPU scheduling all influence how effectively a host can sustain large data flows.

This paper reports on a WLCG mini-capability challenge designed to quantify the impact of host tuning on WAN data transfer performance. The effort brought together multiple ATLAS and CMS sites and used a common tuning approach based on ESnet Fasterdata recommendations and the OSG-provided \texttt{fasterdata-tuning.sh} script, following the established guidance and deployment practices summarized by ESnet and the OSG networking effort~\cite{fasterdata-host-tuning,dtn-tuning,fasterdata-tuning-script}. The goals were to: establish a reproducible tuning workflow; compare baseline and tuned behavior across representative transfer nodes; identify the conditions under which gains are largest; and provide operational guidance for wider deployment.

This work aligns with the broader goal of making host-level optimization a practical and sustainable part of WLCG operations, especially for Data Transfer Nodes (DTNs), perfSONAR test points, and other throughput-oriented services.  

\section{Motivation and challenge context}
\label{sec:motivation}

With the High-Luminosity LHC era approaching\cite{hllhc-tdr} and the WLCG Data Challenge DC27
on the near-term horizon\cite{DC24,dc27-planning}, the pressure to extract maximum performance from
existing infrastructure is acute. Network capacity between major sites has grown
substantially, yet the realized transfer rates at many sites remain well below
what the WAN path can support. The gap frequently originates not in the network
but at the endpoints.

A common experience in large facilities is that transfers are limited by
host-level constraints unrelated to the wide-area path: socket buffer ceilings,
non-optimal queue depths, insufficient NIC ring sizes, offload settings
mismatched to the workload, or storage devices unable to sustain the offered
network rate. A path that is well provisioned at the network layer can still
underperform at the host edge, and the symptoms are easily mistaken for a
network problem.

ESnet's Fasterdata project has codified remedies for these constraints as
deployment guidance for high-throughput hosts~\cite{fasterdata-host-tuning}.
The OSG \texttt{fasterdata-tuning.sh} script operationalizes that guidance
as an auditable, reversible workflow~\cite{fasterdata-tuning-script}. What
has been lacking is systematic evidence of its impact across the variety of
hardware profiles and storage backends present in the WLCG. The mini-capability
challenge reported here was designed to supply that evidence: by applying a
common methodology across multiple ATLAS and CMS sites simultaneously, it
enabled direct site-to-site comparison and exposed the conditions under which
tuning helps, the conditions under which the real bottleneck lies elsewhere,
and the configuration problems that would otherwise have remained invisible.

\section{Methodology}
\subsection{Testbed and participating sites}
\label{sec:testbed}

The challenge comprised two parallel sub-experiments, one organized through
USCMS and one through USATLAS, running during February 2026. EL8/9 hosts with modern NICs at 25\,Gbps or higher were used.
Storage backends varied by site. In each case the baseline and tuned measurements were taken on the
same hardware, which eliminates host-to-host variation as a confound when
interpreting the results. Table~\ref{tab:sites} summarizes the participating
sites.

\begin{table}[h]
\centering
\caption{Participating sites, storage backend, and experiment
assignment. AGLT2 comprises two sub-sites (UM and MSU) with distinct hardware
profiles.}
\label{tab:sites}
\begin{tabular}{@{}llll@{}}
\toprule
Site & Experiment & Tier & Storage \\
\midrule
FNAL       & USCMS   & Tier-1  & dCache (source) \\
UNL        & USCMS   & Tier-2  & XRootD/CephFS   \\
Vanderbilt & USCMS   & Tier-2  & XRootD/LStore\\
UCSD       & USCMS   & Tier-2  & XRootD/CephFS   \\
AGLT2 (UM) & USATLAS & Tier-2  & dCache          \\
AGLT2 (MSU)& USATLAS & Tier-2  & dCache          \\
MWT2       & USATLAS & Tier-2  & dCache         \\
NET2       & USATLAS & Tier-2  & dCache          \\
BNL        & USATLAS & Tier-1  & dCache          \\
\bottomrule
\end{tabular}
\end{table}

{\bf USATLAS:} The USATLAS experiment tested bidirectional production
transfers between AGLT2 (comprising storage at the University of Michigan and
Michigan State University), MWT2 and BNL, using the BNL FTS instance\cite{fts3}. Production
datasets with average file sizes exceeding 4\,GB were used on 2--3 February 2026. Day~1 established
a baseline in the as-built configuration; a dCache\cite{dcache} proxying misconfiguration
was identified during this session and corrected before Day~2, which then
measured the effect of applying the tuning on correctly configured systems.  NET2 was not ready to participate during those 2 days but later ran trans-Atlantic tests with the Prague Tier-2, identifying an ESnet traffic path selection issue that forced transfers onto the shortest path, a single 100 Gbps link, even though both end sites were connected at 400 Gbps; this issue was subsequently reported to ESnet and they fixed the problem.

{\bf USCMS:} FNAL served as the shared data source for all USCMS tests,
contributing 30 of its 150 production dCache servers. The destination sites
were UNL, Vanderbilt, and UCSD. Each destination transferred files from FNAL
using two test types run before and after tuning: multi-threaded iperf3\cite{iperf3}
(32 parallel streams, single node-pair) and production FTS file transfers
with files distributed evenly across all 30 FNAL source nodes available for host tuning tests. The three destination sites were intentionally diverse in NIC speed and storage capability
--- 100\,G DTNs with capable storage (UNL), 25\,G DTNs (Vanderbilt), and newly
deployed 100\,G DTNs connected at 400\,Gbps to ESnet but with storage not yet
matched to that rate (UCSD); enabling a controlled comparison of bottleneck
scenarios under a common test protocol.

\subsection{Tuning approach}

The tuning strategy followed ESnet Fasterdata recommendations and was implemented through the \texttt{fasterdata-tuning.sh} script~\cite{fasterdata-host-tuning,dtn-tuning}.  The script supports two host profiles: \textit{measurement} (perfSONAR nodes)
and \textit{dtn} (storage/transfer nodes), since those systems have different
purposes and may require different tuning.  Originally developed to improve perfSONAR (measurement host) performance, the script was later extended to cover dtn (storage node) profiles.  This automation framework provides a number of practical advantages:

\begin{tightitemize}
  \item audit of current host settings against recommended values;
  \item safe application of tuning parameters using a validated configuration profile;
  \item save, diff, and restore support for testing alternative configurations;
  \item persistence of tuned settings across reboots where appropriate;
  \item operational guidance for packet pacing, offloads, and other host-level features.
\end{tightitemize}

\begin{figure}[htbp]
\centering
\includegraphics[width=0.85\linewidth]{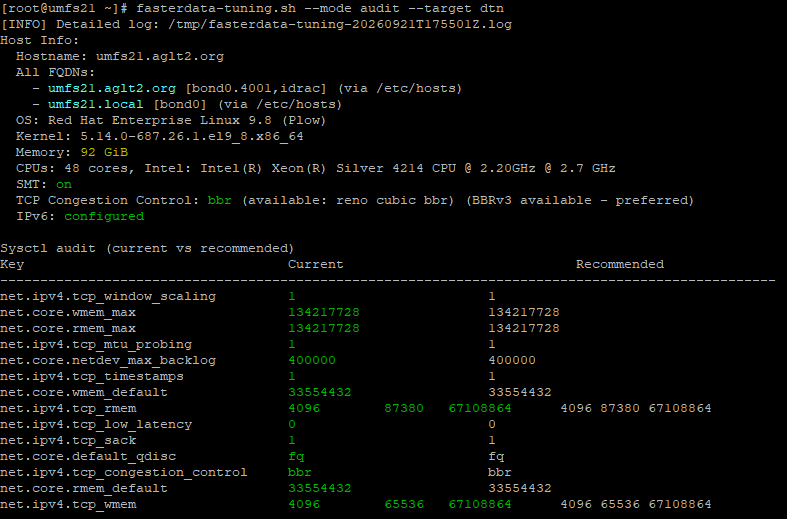}
\caption{Audit-mode output of the \texttt{fasterdata-tuning.sh} script for a storage (dtn) host, showing current versus recommended settings.}
\label{fig:tune-script-ex1}
\end{figure}

\begin{figure}[htbp]
\centering
\includegraphics[width=0.85\linewidth]{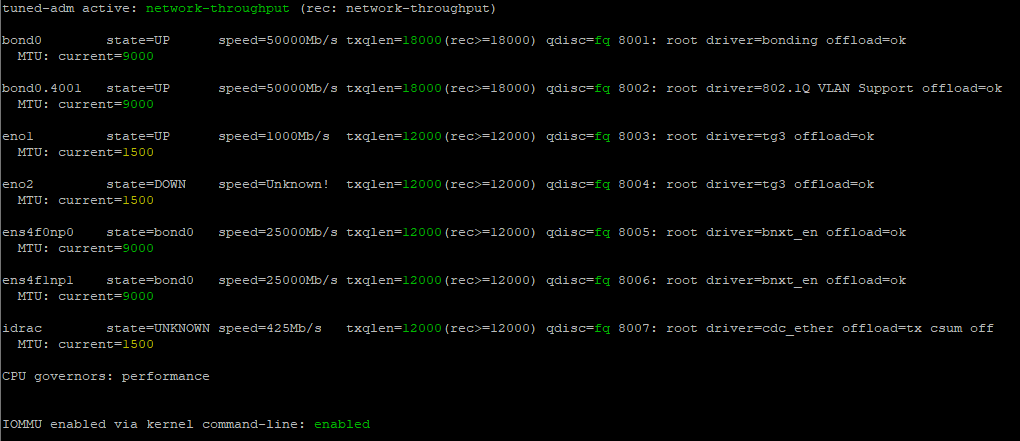}
\caption{NIC analysis output from \texttt{fasterdata-tuning.sh}, showing
interface configuration and recommended settings.}
\label{fig:tune-script-ex2}
\end{figure}

\begin{figure}[htbp]
\centering
\includegraphics[width=0.85\linewidth]{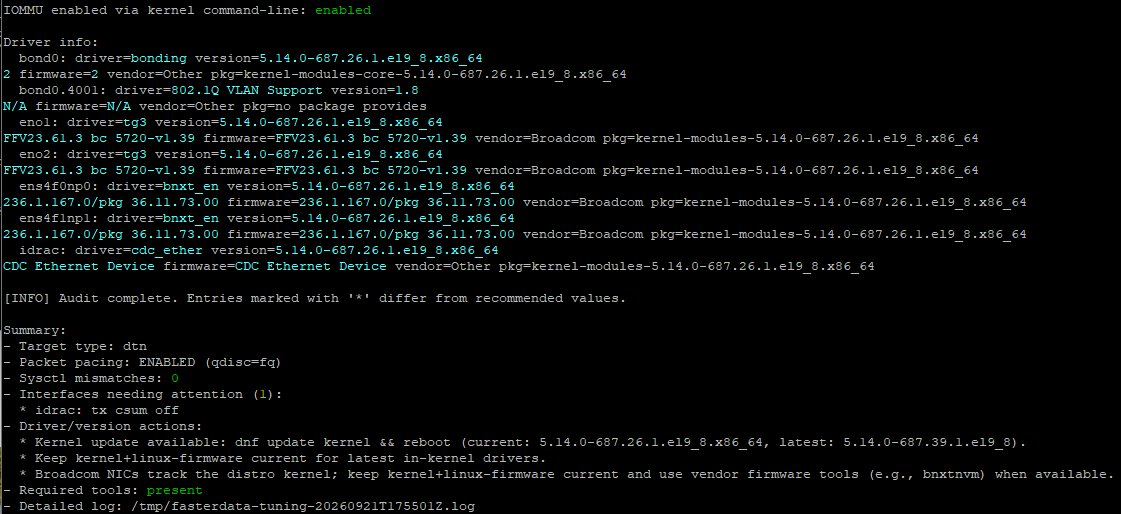}
\caption{Remaining \texttt{fasterdata-tuning.sh} output, including driver
version information and a per-parameter summary.}
\label{fig:tune-script-ex3}
\end{figure}

The tuning areas included parameters commonly associated with high-throughput WAN transfers: TCP socket buffer sizing, queue and backlog settings, interface ring buffer configuration, NIC offload settings, congestion-control selection, I/O scheduler and queue-depth adjustments, and NUMA-aware storage behavior. In DTN-oriented scenarios, we also considered packet pacing and interface-level shaping, since a host can otherwise become bottlenecked by local traffic bursts rather than the remote path itself. Examples of running the tuning script are shown in Figures~\ref{fig:tune-script-ex1}--\ref{fig:tune-script-ex3}.

For reproducibility, the script's save/restore workflow was especially valuable. This allowed the team to preserve a known-good baseline, apply tuning, run transfer tests, and then revert to the original configuration for comparison. This prevented tuning drift between sites and enabled regression checks across host types.

\subsection{Validation and measurement methods}
To evaluate the effect of tuning, the measurement workflow combined host diagnostics with data-transfer tests. The test suite included iperf3 for synthetic host-pair measurements and some sites used fio\cite{fio} for storage-level benchmarking. We then ran representative transfer tests using WLCG-relevant protocols and workflows, including HTTPS Third-Party Copy (HTTPS-TPC) for USCMS and dCache-native FTS transfers for USATLAS.

The principal performance metrics were:

\begin{tightitemize}
  \item total and per-file transfer throughput;
  \item throughput stability across repeated measurements.
\end{tightitemize}

We monitored host CPU and I/O utilization informally to diagnose bottleneck location, but we did not systematically record it across all sites.

Throughput alone does not distinguish network-limited from storage-limited
behavior; site-specific context (NIC speed, storage backend, and path
characteristics) was used alongside the transfer measurements to attribute
observed gains or their absence.

\section{Results and discussion}
\begin{figure}[htbp]
\centering
\includegraphics[width=\linewidth]{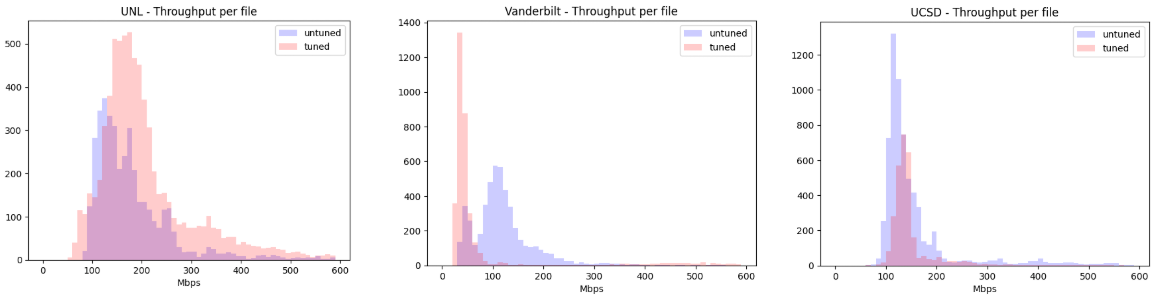}
\caption{FTS transfer throughput before and after applying the dtn tuning
profile at UNL (left, improved), Vanderbilt (center, degraded), and UCSD
(right, approximately unchanged). The UCSD iperf3 result, not shown here,
improved significantly, locating the bottleneck in the storage layer rather
than the network stack.}
\label{fig:host-tuning-results}
\end{figure}

The results show that host tuning for dtn type hosts can meaningfully improve WLCG WAN transfers, but the gains are not uniform across all sites or workloads. In the best cases, tuned hosts achieved substantially higher sustained throughput and lower completion times for otherwise comparable transfers. The largest gains were typically observed where default host settings were clearly sub-optimal for high-rate, long-distance transfers, particularly on systems with modern NICs and sufficiently capable storage backends. Figure~\ref{fig:host-tuning-results} shows the different impacts of host tuning across three representative sites.

At UNL, where both DTNs and storage were capable, FTS throughput rose from
60 to 90\,Gbps following tuning. At UCSD, iperf3 throughput improved from
20 to 90\,Gbps while FTS throughput was unchanged, confirming that the TCP
stack responded to tuning but the storage layer remained the binding constraint.
At Vanderbilt, the 25\,Gbps NIC was already the ceiling, and the dtn tuning
profile reduced throughput, motivating a review of per-site profile selection.

The analysis also highlighted the importance of site-specific tuning. The same configuration change could yield a strong improvement on one host while having limited effect on another, depending on the effective bottleneck. For example, sites limited by default socket tuning or queue depth often recovered a large fraction of their network capability, while sites already close to storage or application limits may even be adversely impacted. This reinforces the importance of auditing each host before applying a broad tuning profile.

A second key result was that the tuning process itself must be operationally safe and reversible. The ability to save the initial system state, apply tuning, compare behavior, and restore the baseline was one of the most important practical features of the workflow. This was essential for controlled comparisons and for reducing operational risk during production use.

The challenge also illustrated the value of integrating tuning with broader monitoring and transfer validation. A host can be tuned correctly but still appear suboptimal if the network path, storage backend, or throughput pattern is not adequately understood. The study therefore treats tuning as one component of a larger operating model for high-performance transfer nodes, rather than as a stand-alone optimization.

\section{Operational implications for WLCG}
The mini-capability challenge experience points to several actionable conclusions for WLCG operations.

First, host tuning should be considered a standard part of DTN\cite{science-dmz} and perfSONAR\cite{perfsonar} deployment, not a one-off site customization. The ESnet Fasterdata guidance provides a strong starting point, and the automation script lowers the barrier to consistent application across sites.

Second, tuning must include a reproducible validation workflow. A host configuration is only useful if you can compare it against a known baseline and restore it when needed. This is particularly important in shared production environments where a tuning change may inadvertently affect unrelated services.

Third, site adoption is most effective when combined with monitoring and documentation. Understanding the network path, NIC capabilities, storage stack, and relevant transfer protocol is essential to interpret performance outcomes and decide whether to roll out a configuration more broadly.

Finally, the challenge suggests that host optimization complements network-level improvements and storage upgrades. It is not a replacement for capacity or path engineering, but it is a cost-effective lever for increasing the efficiency of the existing infrastructure.

\section{Summary and future work}
This WLCG host-tuning mini-capability challenge shows that careful system configuration can improve WAN data-transfer performance in a measurable and reproducible way. The combination of ESnet Fasterdata guidance and the OSG \texttt{fasterdata-tuning.sh} tooling provides a practical path for site administrators to assess and tune production transfer hosts without losing the ability to revert to a known baseline.

This initial exercise was useful for familiarising the team with the tuning script and has shed some light on the complexity of the building blocks involved in a site's throughput capabilities; the lack of consistency in the results calls for a second round of tests, this time with an improved methodology in which different parameters should be tuned and evaluated separately. Storage and network performance should be evaluated both separately and together. 

For some sites (see Figure~\ref{fig:host-tuning-results}), the applied tuning decreased the WAN data-transfer performance.  Since we would like to have the script embody the set of best-practice tuning we can recommend for all sites, it will be critical to identify and fix tunings which worsen data-transfer performance. This will be a primary goal for our next phase of work.

Beyond fixing tunings that decrease performance, the next phase of this effort is to expand the scope beyond a small set of initial hosts and test the operational guidance across a broader WLCG footprint. Additional work is needed to compare tuning outcomes across multiple transfer protocols, storage backends, and path conditions, and to integrate these recommendations more fully into site onboarding and operational procedures.

The overall objective remains clear: to improve the reliability and efficiency of WLCG data movement by aligning host configuration with the realities of modern high-bandwidth scientific workflows.

\section{Acknowledgements}
\enlargethispage*{4mm}
We thank the participating WLCG sites and the broader networking and storage communities for their collaboration in this effort.  We also acknowledge the support of the National Science Foundation through OSG-LHC (OAC-1836650) and IRIS-HEP (PHY-2323298), and the contributions of ESnet, the perfSONAR project, and the LHCONE/LHCOPN communities. Their coordination and operational expertise were essential to complete this work successfully.

%
\bibliography{bibliography}
%
%

\end{document}